\documentclass[useAMS,usenatbib,usegraphicx]{mn2e}
\usepackage{aas_macros}
\usepackage{amsmath,amssymb}

\begin{document}

  \title[Bayesian evidence]{Bayesian evidence: can we beat MultiNest using
  traditional MCMC methods?}
  
  \author[van Haasteren]{Rutger van Haasteren$^1$\footnotemark
  \\
    $^1$Leiden University, Leiden Observatory, P.O. Box 9513, NL-2300 RA
      Leiden, the Netherlands}
 
  \date{printed \today}

  \maketitle

  \begin{abstract}
    Markov Chain Monte Carlo (MCMC) methods have revolutionised Bayesian data
    analysis over the years by making the direct computation of posterior
    probability densities feasible on modern workstations. However, the
    calculation of the prior predictive, the Bayesian evidence, has proved to be
    notoriously difficult with standard techniques. In this work a method is
    presented that lets one calculate the Bayesian evidence using nothing but
    the results from standard MCMC algorithms, like Metropolis-Hastings. This
    new method is compared to other methods like MultiNest, and greatly
    outperforms the latter in several cases. One of the toy problems considered
    in this work is the analysis of mock pulsar timing data, as encountered in
    pulsar timing array projects. This method is expected to be useful as well
    in other problems in astrophysics, cosmology and particle physics.
  \end{abstract}


\footnotetext{Email: haasteren@strw.leidenuniv.nl}

  \section{Introduction}
    Bayesian inference has proved over the years to be a very powerful tool in
    many branches of science as it gives a very clear prescription of how to
    analyse datasets without loss of information, even for very complicated
    models. On the practical side, in performing Bayesian analysis two difficult
    problems often emerge:\newline
    1. Producing the posterior probability density functions (PDFs) for several
    interesting parameters requires marginalisation, i.e.~the integration of the
    full joint posterior PDF over most model parameters. In a majority of cases
    this must be done numerically, and it is common to employ a Markov Chain
    Monte-Carlo (MCMC) algorithm for performing the integration.\newline
    2. Model selection requires the calculation of the Bayes factor: the ratio
    of the prior predictive values of two competing models. This prior
    predictive, or Bayesian evidence as we will call it, is the integral of the
    full joint posterior PDF over all model parameters. The calculation of this
    value can be notoriously difficult using standard techniques.

    The use of Markov Chain Monte Carlo (MCMC) algorithms, such as the
    Metropolis-Hastings algorithm, has become extremely popular in the
    calculation of the marginal posterior PDFs. MCMC algorithms allow
    one to sample from posterior distribution of complicated
    statistical models, greatly reducing the effort involved in evaluating high
    dimensional numerical integrals. The problem with standard MCMC algorithms
    lies in the fact that the normalisation constant of the marginal posterior
    PDFs is lost in the calculation, making it difficult to calculate the
    second of the above mentioned quantities.

    Over the years, several methods capable of calculating the Bayesian evidence
    have been developed, such as using the harmonic mean \citep{Newton1994} and
    parallel tempering \citep{Earl2005}. The problems with these methods are the
    inaccuracy of the method for the harmonic mean, and the high computational
    cost for parallel tempering: this method can easily be 100 times more
    time-consuming than regular MCMC algorithms \citep{Newman1999}. More recent
    efforts include the Nested sampling algorithm \citep{Skilling2004}, a
    greatly improved version of which has been implemented in MultiNest
    \citep[hereafter FHB09]{Feroz2009}.  These recent methods rely on a
    transformation of the integral to a 1-dimensional problem, and then use a
    clever trick for avoiding the calculation of the inverse of the cumulative
    posterior PDF. The MultiNest algorithm of FHB09 has been successfully tested
    on several (toy) problems, and seems to be much more efficient than
    traditional methods.  However, since the MultiNest algorithm by design
    samples from the prior - not from the posterior - we believe that the
    MultiNest algorithm suffers more from the curse of dimensionality than the
    traditional MCMC algorithms \citep{Evans2006}.  When analysing complicated
    models with many parameters ($\sim 100$ or more), the scaling with
    dimensionality is crucial.
    
    In this paper we present a method to calculate the Bayesian evidence from
    the MCMC chains of regular MCMC algorithms. This method can be applied to
    chains that have already been run, provided that the posterior values,
    calculated with normalised prior and normalised likelihood, have been saved
    together with the values of the parameters at each point in the chain. For
    several cases, the accuracy of the new method is significantly greater than
    that of other methods with the same amount of sampled points, provided that
    the MCMC has been properly run. The error on the Bayesian evidence can be
    reliably calculated using a bootstrap procedure.

    The outline of the paper is as follows. In section \ref{sec:bayes} we briefly
    review the basic aspects of Bayesian inference for parameter estimation and
    model selection. Then we provide the reader with some necessary details on
    MCMC in section \ref{sec:MCMC}, where we also outline the new algorithm to
    calculate the Bayesian evidence. In section \ref{sec:comparison} we assess
    the strengths and weaknesses of the new method relative to those that use
    nested sampling or parallel tempering. In section \ref{sec:applications} we
    test all competing algorithms on some toy problems like the analysis of
    pulsar timing data as encountered in pulsar timing array projects.


  \section{Bayesian inference} \label{sec:bayes}
    Bayesian inference methods provide a clear and consistent approach to
    parameter estimation and model selection. Consider a model, or Hypothesis,
    $H$ with parameters $\vec{\Theta}$ for a dataset $\vec{d}$. Then Bayes'
    theorem states that
    \begin{equation}
      P\left( \vec{\Theta}\mid\vec{d}, H\right) =
      \frac{ P\left( \vec{d}\mid\vec{\Theta}, H \right)
      P\left( \vec{\Theta}\mid H \right)}
      { P\left(\vec{d}\mid H \right)},
      \label{eq:bayes}
    \end{equation}
    where $P( \vec{\Theta}) := P( \vec{\Theta}\mid\vec{d}, H)$ is the posterior
    PDF of the parameters, $L(\vec{\Theta}) := P( \vec{d}\mid\vec{\Theta}, H )$
    is the likelihood function, $\pi(\vec{\Theta}) := P( \vec{\Theta}\mid H)$ is
    the prior, and $z := P(\vec{d}\mid H )$ is the prior predictive or Bayesian
    evidence as we call it.

    The Bayesian evidence is the factor required to normalise the posterior over
    $\vec{\Theta}$:
    \begin{equation}
      z = \int L\left(\vec{\Theta}\right)\pi \left(\vec{\Theta}\right)\hbox{d}^{m}\Theta,
      \label{eq:evidence}
    \end{equation}
    where $m$ is the dimensionality of $\vec{\Theta}$. This Bayesian evidence
    can then be used to calculate the so-called odds ratio in favor of model
    $H_1$ over $H_0$, which allows one to perform model selection:
    \begin{equation}
      \frac{P\left( H_1 \mid \vec{d} \right)}{P\left( H_0\mid\vec{d} \right)} =
      \frac{z_1}{z_0}\frac{P\left( H_1 \right)}{P\left( H_0 \right)},
      \label{eq:modelselection}
    \end{equation}
    where $P( H_0 )$ and $P( H_1 )$ are the prior
    probabilities for the different models . As with the prior for
    a model parameters, the prior probability for a model should be chosen to
    reflect the available information.
    
    In traditional MCMC methods as used in parameter estimation problems,
    inferences are obtained by taking samples from a distribution that is
    proportional to the posterior PDF. Therefore one usually ignores the
    normalisation constant.
    This is acceptable since these methods lose the information about the
    normalisation anyway.  But in contrast to parameter estimation problems, in
    model selection the Bayesian evidence plays a central role: it is a measure
    for how well the data supports the model.
    
    The average of the likelihood over the prior distribution, the evidence, is
    larger for a model if more of its parameter space is likely and smaller for
    a model with large areas in its parameter space having low likelihood
    values. Even if the likelihood function has high peaks, in order to increase
    the evidence these peaks must compensate for the areas in its parameter
    space where the likelihood is low. Thus the evidence automatically
    implements Occam's razor: a simpler theory with compact parameter space will
    have a larger evidence than a more complicated model, unless the latter is
    significantly better at explaining the data.

  \section{Markov Chain Monte Carlo} \label{sec:MCMC}
    Markov Chain Monte Carlo methods (MCMC) can be used to sample from very
    complicated, high dimensional distributions; for Bayesian inference it is the
    posterior, but it could be any distribution. The method presented in this paper
    could be useful for integration problems other than Bayesian evidence
    calculation, so we use the more general $f( \vec{\Theta })$ to denote this
    function. The samples drawn from this function can then be used to perform
    the integrals we need for the marginal posterior PDFs and, as we show,
    for the Bayesian evidence. The idea is quite straightforward: a large number
    of samples drawn from a distribution proportional to $f( \vec{ \Theta })$
    will end up being distributed with a sample density proportional to $f(
    \vec{ \Theta} )$. The exact mechanism that produces these samples can differ
    between MCMC methods and is irrelevant for the purposes of this paper, but
    the result is always a large number of samples distributed according to $f(
    \vec{ \Theta })$. The main advantage of this is that we do not have to
    sample from the entire volume of the parameter space or the prior, but only
    from a small fraction of it: the fraction where the function has a high
    value. Especially for functions in high-dimensional parameter spaces this
    feature is crucial.

    In section \ref{sec:volume} we show that all information needed to
    calculate the Bayesian evidence is already present in the samples of the
    MCMC. Then we give a practical example of how to calculate the integral
    in practice in section \ref{sec:practicalalgorithm}.

    \subsection{Volume calculation} \label{sec:volume}
\begin{figure}
	\includegraphics[width=0.5\textwidth]{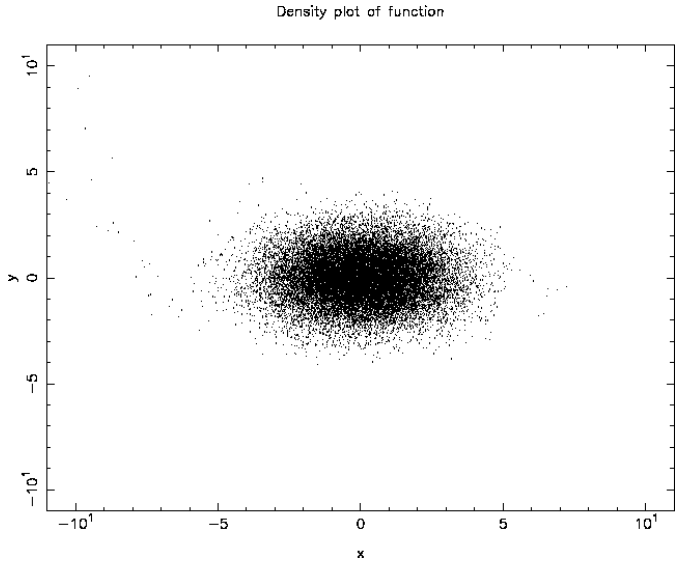}
	\caption{A scatter plot of of $40000$ samples, drawn using a Metropolis
	  algorithm from the function $f( x, y ) = \exp (
	  -ax^2-by^2 )$, with $a=1/5$ and $b=2/5$. We have included the
	  burn-in period in this plot, even though that part of the chain should
	  be discarded for any further calculations.}
	\label{fig:gaussamples}
      \end{figure}

      Consider the unnormalised distribution:
      \begin{equation}
	f\left( x, y \right) = \exp\left( -ax^2-by^2 \right),
	\label{eq:gaussian2dpdf}
      \end{equation}
      where $a$ and $b$ are arbitrary model parameters. For the values
      $a=1/5$ and $b=2/5$, a Metropolis algorithm with $N=40000$ samples 
      yield a distribution of samples similar to Figure \ref{fig:gaussamples}.
      We use the notation $\vec{\Theta}_i = (x_i, y_i)$ to indicate the
      $i^{\hbox{th}}$ sample. We would now like to calculate the integral
      \begin{equation}
	I = \int\int f\left( x, y \right) \hbox{d}x\hbox{d}y .
	\label{eq:gaussian2dintegral}
      \end{equation}
      \begin{figure}
	\includegraphics[width=0.5\textwidth]{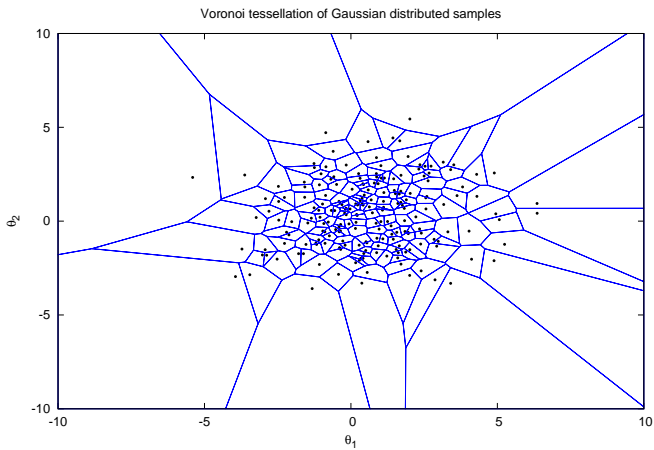}
	\caption{An example of a Voronoi tessellation. Here we have taken
	  $200$ samples from a $2$-dimensional Gaussian distribution as the
	  centres of the Voronoi diagram.}
	\label{fig:voronoi}
      \end{figure}
      In MCMC algorithms, it is usually required or straightforward to calculate
      the function values $f( \vec{\Theta}_i)$ for each sample, and
      these values are often stored together with the
      values of the parameters. These function values can be used to calculate
      the integral if we treat the MCMC samples in parameter space as an
      irregular grid. The most obvious and exact way to do this is to calculate
      the Voronoi tessellation for the samples in the parameter space, an
      example of which is given in Figure \ref{fig:voronoi}. The samples are
      then the centres of the Voronoi cells, and the integral can be calculated
      as follows:
      \begin{equation}
	I \approx \sum_{i} f\left( \vec{\Theta}_i \right) O_i,
	\label{eq:voronoiintegral}
      \end{equation}
      where $O_i$ is the area of the $i^{\hbox{th}}$ Voronoi cell, and we
      only sum over closed cells (with finite area). Theoretically, this
      procedure converges to the correct value in all cases for large number of
      samples $N$. However, although Voronoi tessellations can be computed for
      any dimensionality, this becomes computationally prohibitive in practice
      for problems with high dimensionality \citep{Edelsbrunner1992}. This
      procedure does illustrate that all information needed to evaluate the
      integral $I$ is present in the MCMC chain.

    \subsection{Estimating the covered parameter volume} \label{sec:parametervolume}
      As we have shown in section \ref{sec:volume}, the integral $I$ can be
      approximated by the summation over the function values $f(
      \vec{\Theta}_i)$ times the Voronoi cell size $O_i$. The value $O_i$ is the
      parameter volume occupied by the Voronoi cell of the $i^{\hbox{th}}$ MCMC
      sample, so in the rest of this paper we call $O_i$ the parameter volume of
      the $i^{\hbox{th}}$ sample. Similarly, the parameter volume of the entire
      chain can now be approximated by:
      \begin{equation}
	O = \sum_{i=1}^{N} O_i,
	\label{eq:parametervolume}
      \end{equation}
      where $O$ is the parameter volume of the entire MCMC chain, and $N$ is the
      number of samples in the chain.

      We now show that we do not need the Voronoi tessellations to
      calculate the parameter volume and the integral. For the rest of this
      section we regard the more general case that we have an
      $n$-dimensional parameter space, instead of the $2$-dimensional case of
      the above example. In any MCMC algorithm, the density of points of the
      chain at a certain point $\vec{\Theta}_i$ in parameter space will become proportional
      to the function value $f( \vec{\Theta}_i )$ for large $N$. Therefore,
      \begin{equation}
	\lim_{N\to \infty} O_i = \frac{\alpha}{f\left( \vec{\Theta}_i \right)},
	\label{eq:proportionalparametervolume}
      \end{equation}
      where $\alpha$ is a proportionality constant to be determined. If we somehow
      can estimate the constant $\alpha$ for this chain, we can immediately calculate
      the integral by $I = N\alpha$.

      The proportionality constant $\alpha$ can be estimated for a certain chain by
      regarding a small subset $F_{\hbox{t}} \subset \mathbb{R}^n$ of the
      parameter space for which:\newline
      1. we can calculate what the volume $V_{\hbox{t}}$ of the small
      subset
      is.\newline
      2. the subset $F_{\hbox{t}}$ is sufficiently and completely populated by MCMC
      samples: all parts of $F_{\hbox{t}}$ should have a sample density
      proportional to $f( \vec{\Theta} )$\newline
      MCMC algorithms are employed in order to avoid having to draw samples from
      the entire parameter space, so we have to be careful to choose the right
      test subset in order to fulfill requirement 2. If we have found such a
      subset $F_{\hbox{t}}$, the following equation allows us to estimate $\alpha$:
      \begin{equation}
	V_{\hbox{t}} = \sum_{\vec{\Theta}_i \in F_{\hbox{t}}}
	\frac{\alpha}{f\left( \vec{\Theta}_i
	\right)}.
	\label{eq:proportionalconstant}
      \end{equation}
      The calculation of the integral is now trivial:
      \begin{equation}
	I = N\alpha.
	\label{eq:Nalpha}
      \end{equation}

    \subsection{A practical algorithm} \label{sec:practicalalgorithm}
      In this section we construct a simple practical algorithm for numerically
      calculating integrals using regular MCMC methods. As stated in section
      \ref{sec:parametervolume}, we need to define a small subset of the
      parameter space that is sufficiently populated by MCMC samples. The most
      obvious choice would be to look for a peak in the distribution. In the
      case of multimodal posteriors, we should take samples that 'belong' to one
      specific mode. Clustering algorithms like X-means \citep{Pelleg2000}, and
      G-means \citep{Hamerly2003} can be used for this.

      Assuming that we have found $M$ samples that belong to a specific mode, we
      can then define a subset $F_{\hbox{t}}$ as follows. Assume that the $M$ samples are
      sorted according to their function value $f( \vec{\Theta}_i )$,
      with $f( \vec{\Theta}_0 )$ the highest value. We then first
      approximate the maximum of $f( \vec{\Theta} )$ as:
      \begin{equation}
	\vec{\mu} = \frac{1}{k}\sum_{i=1}^{k} \vec{\Theta}_i,
	\label{eq:mean}
      \end{equation}
      where $k=aM$ is a small fraction of $M$, dependent on the shape of the
      mode.  We use $a = 1/20$ in several cases. We now define our subset
      $F$ as an ellipsoid with centre $\vec{\mu}$, and covariance matrix $C$,
      defined by:
      \begin{equation}
	C = \frac{1}{n} \sum_{i=1}^{n} \left( \vec{\Theta}_i - \vec{\mu}
	\right)\left( \vec{\Theta}_i - \vec{\mu} \right)^T,
	\label{eq:covariance}
      \end{equation}
      where $n=bM$ is a fraction of $M$, again dependent on the shape of the
      mode.  We use $b = 1/5$ in several cases. All samples within this
      ellipsoid of size $\sqrt{r^2\det C}$ satisfy:
      \begin{equation}
	\left(\vec{\Theta} - \vec{\mu}\right)^{T} C^{-1} \left(\vec{\Theta} -
	\vec{\mu}\right) \leq r^2.
	\label{eq:ellipsoid}
      \end{equation}
      We adjust the parameter $r^2$ such that $l=cM$ samples satisfy
      this relation. It is crucial that we choose $l$ to be as large as
      possible, while still satisfying the requirement that the entire ellipsoid
      is sufficiently populated with samples. If $l$ is too small, we will not
      achieve very high accuracy for our integral $I$, but if $l$ is too large,
      we have underpopulated parts of the ellipsoid which results in
      the wrong outcome. We use $c=1/3$ in several cases.

      We now have all the ingredients needed to calculate the integral
      $I$, as the Volume of our $k$-dimensional subset is given by:
      \begin{equation}
	V_{\hbox{t}} = \frac{r^{k}\pi^{\frac{k}{2}}}{\Gamma\left(
	1+\frac{k}{2} \right)} \sqrt{\det C}.
	\label{eq:ellipsoidvolume}
      \end{equation}
      This, combined with equations (\ref{eq:proportionalconstant}) and
      (\ref{eq:Nalpha}) allows us to evaluate the integral.

      We would like to note that this prescription is merely one of many that
      could be used. In this case we have defined our subset $F_{\hbox{t}}$ as an
      ellipsoid located at a maximum of our integrand, but any other subset that
      meets the criteria of section \ref{sec:parametervolume} will suffice.

    \subsection{Error estimates}
      A MCMC algorithm generates samples according to the
      distribution $f(\vec{\Theta})$. Consider the amount of samples inside the
      subset $F_{\hbox{t}}$. This amount follows a Poissonian
      distribution with mean and variance equal to $cM$. We thus obtain the
      theoretical estimate for the error in the integral:
      \begin{equation}
	\Delta I = \frac{1}{\sqrt{cM}}I.
	\label{eq:integralerror}
      \end{equation}
      As we show in section
      \ref{sec:highdimensional}, this estimate is reliable for Monte Carlo algorithms
      that yield uncorrelated samples. In practice, however, many MCMC algorithms produce
      correlated samples since a whole
      chain of samples is used \citep{Roberts1997}.
      As we show in section \ref{sec:applications}, the use of
      correlated MCMC samples for numerical integral as proposed in this paper
      invalidates the error estimate of
      equation~(\ref{eq:integralerror}).

      Having efficiency as one of our goals, we would like to produce reliable
      error-bars on our numerical integral using the correlated MCMC samples. We
      propose to use a bootstrap method \citep{Efron1979}. In the case of several
      chains, one can use the spread in the estimates based on different chains
      to estimate the error of the integral. In the case of a single chain, we
      propose to divide the chain in $10$ succeeding parts, and use the spread
      of the integral in those $10$ parts to estimate the error of the numerical
      integral. In section \ref{sec:errortests} we test the error estimates
      discussed here extensively.

  \section{Comparison to other methods} \label{sec:comparison}
    Although the formalism developed in this paper is quite generally
    applicable, in practice there are caveats for each integration
    problem that one needs to be aware of in order to choose the right integration
    algorithm. In this section we discuss the strengths and the weaknesses
    of the method developed in this paper, and we compare it to 
    Nested sampling, and Parallel tempering.

    For all algorithms, the main criteria that we cover are efficiency, accuracy
    and robustness.

    \subsection{Method using traditional MCMC}
      The main problem that MCMC algorithms try to tackle is that of parameter
      space reduction; the full parameter space over which we would like to
      evaluate the integral is too large, which is why we cannot use a regular
      grid. The idea behind MCMC algorithms is that of \emph{importance
      sampling}: only draw samples in the regions of parameter space where the
      function value $f( \vec{\Theta} )$ is high enough to
      significantly contribute to the integral. This can be done by drawing
      samples from a distribution $P( \vec{\Theta} )$ that resembles
      the function $f( \vec{\Theta} )$, with the optimal choice being
      $P( \vec{\Theta} ) = f( \vec{\Theta} )$. MCMC are
      specifically designed to satisfy this optimal relation, making MCMC
      methods optimally efficient from a theoretical point of view
      \citep{Roberts1997}.
      In Figure \ref{fig:gaussamples} we show an example of samples generated
      with a Metropolis MCMC algorithm, released on a $2$-dimensional Gaussian
      function.

      The drawback of MCMC algorithms is that due to the parameter space
      reduction one can never be sure that all important parts of the parameter
      space have been covered. Some functions have many distinct peaks, the
      so-called modes of the function. MCMC algorithms often have difficulty to
      make the chain move from one mode to another.  If the function $f(
      \vec{\Theta} )$ is highly multimodal, we must make sure that the sample
      density ratio between the modes is right and that we have reached all the
      important modes in the entire parameter space.

      An additional practical challenge with the method developed in this paper
      is to make sure that we have constructed a subset of the parameter
      space that is sufficiently populated with samples. In the case of a not
      very peaked, or highly degenerate function this requires special care.

    \subsection{Nested sampling}
      The nested sampling algorithm is a Monte Carlo technique aimed at accurate
      evaluation of numerical integrals, while staying efficient
      \citep{Skilling2004}.
      The algorithm solves the problems of regular MCMC algorithms by starting
      with sampling from the original parameter space. In the case of Bayesian
      evidence calculation this is equivalent to sampling from the prior
      distribution of the parameters. The density with which the parameter space
      is sampled is adjustable in the form of an amount $n_L$ of so-called live points;
      the number of points evenly distributed among the part of the parameter
      space we are exploring. At the start of the algorithm, the live points are
      evenly distributed over the entire parameter space.
      Then parameter space reduction is achieved 
      by replacing the live points one-by-one under the restriction
      that the newly sampled live points are higher than the lowest one we
      have not replaced yet. Effectively this is the same as shrinking the
      parameter space by a fixed factor every time we replace a new live point,
      ultimately sampling only the part of the function close to the maximum.

      A big advantage of the nested sampling algorithm is that one generates
      samples directly from the whole parameter space. If $n_L$ is chosen high
      enough, i.e.~there is a high enough sampling density in the parameter
      space, there is a very low probability that a mode of the distribution
      $f(\vec{\Theta})$ will be missed. This is one of the difficulties in many
      MCMC implementations.
      
\begin{figure}
      \includegraphics[width=0.5\textwidth]{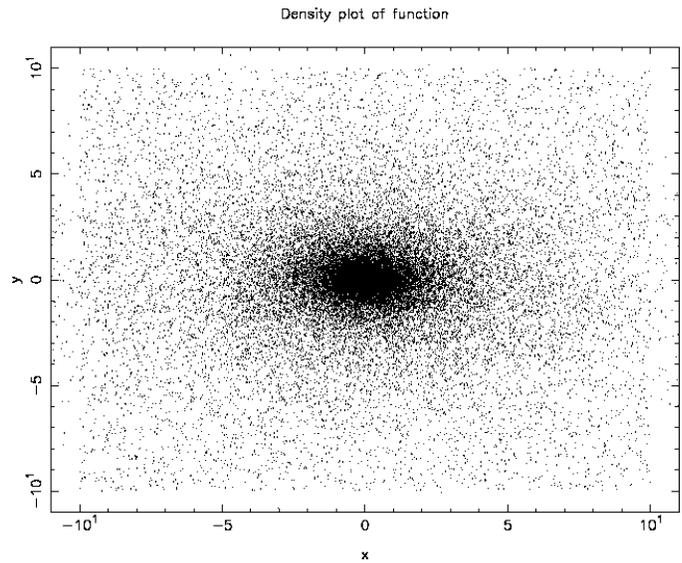}
      \caption{A scatter plot of $40000$ samples, drawn using a nested sampling
	algorithm with $5000$ live points from the function $f( x, y ) = \exp (
	-ax^2-by^2 )$, with $a=1/5$ and $b=2/5$.}
      \label{fig:multinestsamples}
    \end{figure}

      A big disadvantage of the nested sampling algorithm is that one directly
      generates samples from the whole parameter space, instead of from a
      distribution that more closely resembles the function $f(
      \vec{\Theta} )$ we want to integrate. This method will therefore
      never reach the efficiency that traditional MCMC methods offer. In Figure
      \ref{fig:multinestsamples}  we show an example of samples generated with a
      nested sampling algorithm, applied to a $2$-dimensional Gaussian function.
      In higher dimensional problems, as discussed in section
      \ref{sec:highdimensional}, the direct sampling from the whole parameter
      space can lead to serious efficiency problems that need to be solved.

    \subsection{Parallel tempering}
      Parallel tempering (PT) is an algorithm spawned from the desire to solve
      the problems of traditional MCMC methods. PT algorithms possess better
      mixing properties, allowing the chain to ``escape'' local extrema, and
      allow one to calculate the complete integral, or in our case the Bayesian
      evidence \citep{Earl2005}.  Let us briefly review PT in this section,
      without discussing it in too much detail.

      The main idea of PT is that of parameter space exploration by adding an
      imaginary inverse temperature $\beta$ to the system, changing the
      integrand of our integral to:
      \begin{equation}
	f_{\beta}\left( \vec{\Theta} \right) = \left( f( \vec{\Theta}
	) \right)^\beta.
	\label{eq:ptfunction}
      \end{equation}
      Then many MCMC chains are released in the parameter space, each with a
      different temperature $\beta \in \left[ 0, 1 \right]$. A clever
      swapping system is employed, allowing the chain with $\beta=1$ - the
      ``true'' chain - to swap parameters with chains of higher temperature every
      now and then provided that such a high-temperature chain was able to reach
      a point in parameter space with high $f( \vec{\Theta} )$. This trick
      allows the coldest system, the ``true'' chain with $\beta=1$, to escape
      local extrema.

      The integral over $f(\vec{\Theta})$ is calculated by using all chains,
      not just the one with $\beta=1$, as follows. We first define a partition
      function:
      \begin{equation}
	Z(\beta) = \int \hbox{d}\vec{\Theta} f_{\beta}\left(\vec{\Theta}\right),
	\label{eq:partition}
      \end{equation}
      which has a logarithmic derivative of:
      \begin{equation}
	\frac{\hbox{d}}{\hbox{d}\beta} \log\left( Z(\beta) \right) =
	\frac{1}{Z(\beta)} \frac{\hbox{d}}{\hbox{d}\beta} Z(\beta) =
	\left\langle \log\left( f(\vec{\Theta}) \right) \right\rangle_\beta,
	\label{eq:partitionderivative}
      \end{equation}
      where $\langle.\rangle_\beta$ is the expectation value of a quantity over
      a distribution proportional to $f_{\beta}(\vec{\Theta})$. Since we know
      that our desired integral can be expressed as $I = Z(1)$, equation
      (\ref{eq:partitionderivative}) is all we need to calculate it:
      \begin{equation}
	\log\left(Z(1)\right) = \log\left( Z(0) \right) +
	\int_{0}^{1}\hbox{d}\beta \left\langle \log\left( f(\vec{\Theta})
	\right) \right\rangle_{\beta}.
	\label{eq:integral}
      \end{equation}
      The observant reader will have noted that we have neglected to mention the
      size $Z(0)$ of the parameter space that we explore with our chains. The
      high temperature chain with $\beta=0$ is unbounded by the function
      $f(\vec{\Theta})$, and therefore will transverse the entire parameter
      space. We should make sure that we limit the size of the parameter space
      as much as possible, without missing any peaks of $f(\vec{\Theta})$.

      The main advantages of Parallel tempering are that it explores the entire
      parameter space, even in the presence of strong local peaks, and that the
      Bayesian evidence can be calculated. These advantages are accompanied with
      the large computational costs of the extra chains with $\beta \neq 1$,
      which has resulted in the need for alternative methods like MultiNest. As
      MultiNest is supposed to outperform PT in virtually all cases (FHB09), we
      will compare our method to MultiNest only in section
      \ref{sec:applications}.

  \section{Applications and tests} \label{sec:applications}
    In this section we consider several toy models, and we apply
    several integration algorithms to these toy models as to compare the
    algorithms. In this whole section we have two ways to use the MCMC algorithm.
    To produce correlated samples, we use a regular Metropolis-Hastings
    algorithm where we use the entire chain. To produce uncorrelated samples, we
    use the same algorithm, but we only store one in every $j$ samples
    produced by the algorithm. This does require us then to run the chain
    $j$ times longer to produce the same number of samples. We choose $j$ high enough that there is
    negligible correlation between 2 succeeding used samples; in the case of a
    $n$-dimensional Gaussian as in section \ref{sec:highdimensional} we use
    $j=100$.

    \subsection{Toy model 1: a high-dimensional Gaussian} \label{sec:highdimensional}
      We first consider a problem that is a typical example of what MCMC
      algorithms where designed for: a highly peaked, high-dimensional function.
      The curse of dimensionality prohibits any direct numerical integration
      scheme on a fine grid, but analytical integration is possible. Consider
      the multiplication of $n$ Gaussian functions,
      \begin{equation}
	f_{1}\left( \vec{x} \right) = \prod_{i=1}^{n}
	\sqrt{\frac{a_{i}}{2\pi}}\exp\left( -\frac{1}{2}a_{i}x_{i}^{2} \right),
	\label{eq:gaussianmultiplication}
      \end{equation}
      with $a_i$ the width of the Gaussian in the $i^{\hbox{th}}$ direction. Now
      let us perform a volume preserving coordinate transformation using a
      random orthogonal matrix $R$ as follows: $\vec{\Theta} = R\vec{x}$. If we
      introduce a matrix $A^{-1}_{ii} = a_i = 1+i$ with $A^{-1}_{ij} = 0$ for $i
      \neq j$, we then have a highly degenerate high-dimensional Gaussian
      function:
      \begin{equation}
	f_{1}\left( \vec{\Theta} \right) = \frac{1}{\left( 2\pi
	\right)^{n/2}\sqrt{\det C}} \exp \left( -\frac{1}{2}\vec{\Theta}^{T}
	C^{-1} \vec{\Theta} \right),
	\label{eq:gaussianmultidimensional}
      \end{equation}
      where we have introduced the coherence matrix $C = R A R^T$.

      We now apply in turn the MultiNest algorithm and the algorithm developed in
      this paper combined with Metropolis-Hastings to the multi-dimensional
      Gaussian for various number of dimensions.
      
      For the MultiNest algorithm, we use a number of live points $L = 50n$ with
      $n$ the number of dimensions, and we have set the sampling efficiency to
      $e = 0.3$ and the evidence tolerance to $t=0.1$ as advocated in
      FHB09\footnote{We have not used wrapping for the parameters. Wrapping does
      speed up the sampling, but somehow te resulting evidence estimates are
      less accurate}.
      
      We set up the Metropolis-Hastings algorithm to have an acceptance
      ratio equal to the optimal value of $23.4\%$ \citep{Roberts1997}. The parameters
      of the algorithm advocated in section \ref{sec:practicalalgorithm} have
      been set to: $a = 1/20, b = 1/5, c = 1/3$. We have used the number of
      samples $N$ used by the MultiNest algorithm as the number of samples in our
      MCMC chain. However, we realise that the MultiNest algorithm might be
      improved in the future, as the algorithm is still under construction. The
      lowest efficiency (used samples / drawn samples) we encountered for this
      toy-problem was $e = 0.08$; we therefore estimate that the error-bars
      could be decreased with a factor of maximally $\sqrt{1 / 0.08} \approx
      3.5$. The results of this toy-model are shown in table \ref{tab:toy1}.
      \begin{table}
	\centering
	\begin{tabular}{r r r r r r r}
	  \multicolumn{5}{c}{$\log\left( I \right)$ for different algorithms} \\
	  \hline
	  n & \# N & MultiNest & Unc. MCMC & Cor. MCMC \\
	  \hline
	  $2$ & $2902$     & $-0.17 \pm 0.18$ & $-0.018 \pm 0.025 $ & $ 0.03  \pm 0.025 $ \\
	  $4$ & $7359$     & $0.20 \pm 0.17$  & $ 0.007 \pm 0.024 $ & $-0.01  \pm 0.03  $ \\
	  $8$ & $24540$    & $0.17 \pm 0.17$  & $-0.01  \pm 0.01  $ & $ 0.02  \pm 0.03  $ \\
	  $16$ & $10^5$    & $0.05 \pm 0.18$  & $ 0.001 \pm 0.006 $ & $ 0.004 \pm 0.03  $ \\
	  $32$ & $10^6$    & $1.43 \pm 0.17$  & $ 0.004 \pm 0.004 $ & $-0.015 \pm 0.010 $ \\
	  $64$ & $4.10^6$  &                  & $-0.0004 \pm 0.0007$ & $-0.02 \pm 0.016 $ \\
	  $128$ & $10^7$   &                  & $0.0006 \pm 0.0004$ & $-0.05  \pm 0.03  $ \\
	  \hline
	\end{tabular}
	\caption{The log-integral values of the function $f_{1}$ of
	  equation (\ref{eq:gaussianmultidimensional}). $N$ is the number of
	  samples, and $n$ is the number of dimensions.
	  The analytically integrated
	  value is $\log{I} = 0$ for all values of $n$. For $n \geq 64$, we were
	  not able to successfully complete the MultiNest algorithm with said
	  parameters due to limited computational power. Memory size was the
	  limiting factor for the regular MCMC.}
	\label{tab:toy1}
      \end{table}

    \subsection{Toy model 2: egg-box function} \label{sec:eggbox}
\begin{figure}
	\includegraphics[width=0.5\textwidth]{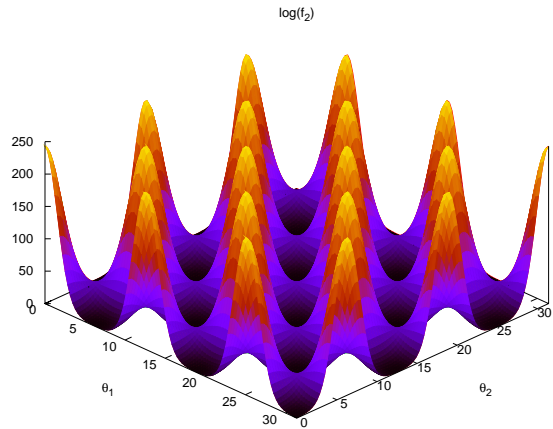}
	\caption{Toy-model 2: the eggbox of equation~(\ref{eq:eggbox})}
	\label{fig:eggbox}
      \end{figure}

      Just as in FHB09, we now consider a highly multimodal two-dimensional
      problem for which the function resembles an egg-box. The function is
      defined as:
      \begin{equation}
	f_{2}\left( \vec{\Theta} \right) = \exp\left[ 2 + \cos\left( \Theta_1
	\right)\cos\left( \Theta_2 \right) \right]^5,
	\label{eq:eggbox}
      \end{equation}
      where we set the domain of the function equal to $[ 0, 10\pi
      ]$ for both parameters. The shape of this function is shown in
      Figure \ref{fig:eggbox}. This is a typical problem where difficulties
      arise for traditional MCMC algorithms. Many solutions have been proposed
      for situations like this \citep{Newman1999}, but in practice one needs to
      have additional information about the problem for any of those solutions to be
      reliable. For
      the sake of clarity of this paper, we do not concern us with the
      practical implementation of the MCMC algorithm. We assume that a
      suitable trick can be found for the problem at hand so that the
      algorithm proposed in this paper can be used. For the eggbox toy-model we
      will use a jump-technique. At each iteration of the MCMC algorithm, there
      is a small probability, $1\%$ in this case, that the chain will jump to a
      random neighbouring mode.

      The MultiNest algorithm is ideally suited for this problem, as this is a
      low-dimensional multimodal function. With enough live samples all modes
      should be easily discovered, and the peaks are symmetric and well-behaved.
      We run MultiNest on this problem with the same parameters as in FHB09: We
      use $L = 2000$ live points, efficiency $e = 0.3$, and tolerance $t = 0.1$.

      Table \ref{tab:eggbox} shows the results of this analysis\footnote{The
      true value is different from the one noted in FHB09. We have taken the
      hypercube transformation into account in our calculations.}. For
      evaluating the integral with the MCMC chain, we have taken a total of
      $N=60210$ samples as was done with MultiNest, but we have used only the
      samples of one single peak in equation (\ref{eq:proportionalconstant}). The
      amount of samples in a single peak is $2/25$ of the total amount of
      samples, leading to loss of accuracy. Though more sophisticated methods
      can be constructed by, say, averaging the values of the proportionality
      constant $\alpha$ of equation (\ref{eq:proportionalparametervolume})
      for all individual peaks, we show here that we only need to find one
      single portion of the parameter space that is sufficiently populated to
      calculate a reliable value for the integral.

      \begin{table}
	\centering
	\begin{tabular}{r r r r}
	  \multicolumn{4}{c}{$\log\left( I \right)$ for different algorithms} \\
	  \hline
	  \# N & MultiNest & Unc. MCMC & Cor. MCMC \\
	  \hline
	  $60210$  & $240.19 \pm 0.05$ & $240.19 \pm 0.027$   & $240.23 \pm 0.05$ \\
	  \hline
	\end{tabular}
	\caption{The log-integral values of a single mode of the eggbox function
	  of equation~(\ref{eq:eggbox})
	  The fine-grid
	  integrated value is $\log{I} = 240.224$}
	\label{tab:eggbox}
      \end{table}

    \subsection{Application in Pulsar Timing} \label{sec:pulsartiming}
      In pulsar timing data analysis, one often encounters datasets of which the
      exact statistics are not well known. Bayesian model selection would
      provide the ideal tool to obtain information about the power spectra
      present in the data. \citet[hereafter vHLML]{vanhaasteren2009} give a full
      description of the Bayesian data analysis for pulsar timing arrays, but
      their work lacks a method to calculate the Bayesian evidence from the MCMC
      samples. In this section, we use a pulsar timing mock dataset to show that
      the method developed in this paper is well-suited for Bayesian evidence
      calculation in pulsar timing problems. We also use MultiNest to analyse
      this dataset, and we compare the two results.

      For the purposes of this paper, the description of the pulsar timing
      posterior distribution is kept brief; full details can be found in vHLML.
      The data of pulsar timing experiments consists of the arrival times of
      pulsar pulses (TOAs), which arrive at the earth at highly predictable
      moments in time \citep{Hobbs2006}. The deviations from the theoretically
      predicted values of these TOAs are called the timing-residuals (TRs).
      These TRs are the data we concern ourselves with in this example.

      Consider $n=100$ TRs, denoted as $\vec{\delta t}$, observed
      with intervals between succeeding observations of $5$ weeks. Assume
      that we are observing a stable and precise millisecond pulsar with timing
      accuracy about $\sigma = 100$ns (the error bars on the TRs).
      Usually $\sigma$ not precisely known, since pulsar timers generally assume
      that their estimate of the error bar is slightly off. Several datasets of
      millisecond pulsars also seem to contain correlated low frequency noise
      \citep{Verbiest2009}. We therefore also allow for some correlated
      timing-noise in the data, with a power-spectrum given by:
      \begin{equation}
	S(f) = r^2\gamma\exp\left( -\gamma f \right),
	\label{eq:powerspectrum}
      \end{equation}
      where $f$ is the frequency, $r$ is the amplitude of the correlated timing-noise in ns, and $\gamma$ is
      the typical size of the structures that appear in the data due to this
      correlated timing-noise. Following vHLML, we can now write the likelihood
      function for the TRs as a multi-dimensional Gaussian:
      \begin{equation}
	P\left(\vec{\delta t} \mid \sigma, r, \gamma\right) =
	\frac{1}{\sqrt{\left( 2\pi \right)^n \det C}} \exp\left(
	-\frac{1}{2}\vec{\delta t}^{T} C^{-1} \vec{\delta t}
	\right),
	\label{eq:ptagaussian}
      \end{equation}
      where $C$ is an $(n\times x)$ matrix, with elements defined as:
      \begin{equation}
	C_{ij} = \sigma^2\delta_{ij} + r^2\frac{\gamma^2}{\gamma^2+\tau_{ij}^2},
	\label{eq:coherence}
      \end{equation}
      with $\delta_{ij}$ the Kronecker delta, and $\tau_{ij}$ is the time
      difference between observation $i$ and observation $j$.

\begin{figure}
	\includegraphics[width=0.5\textwidth]{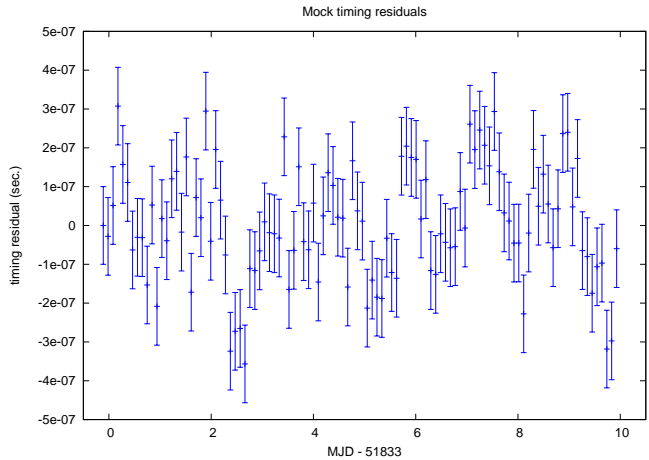}
	\caption{The mock timing-residuals we have analysed with both MultiNest
	  and the method developed in this paper}
	\label{fig:residuals}
      \end{figure}

\begin{figure}
	\includegraphics[width=0.5\textwidth]{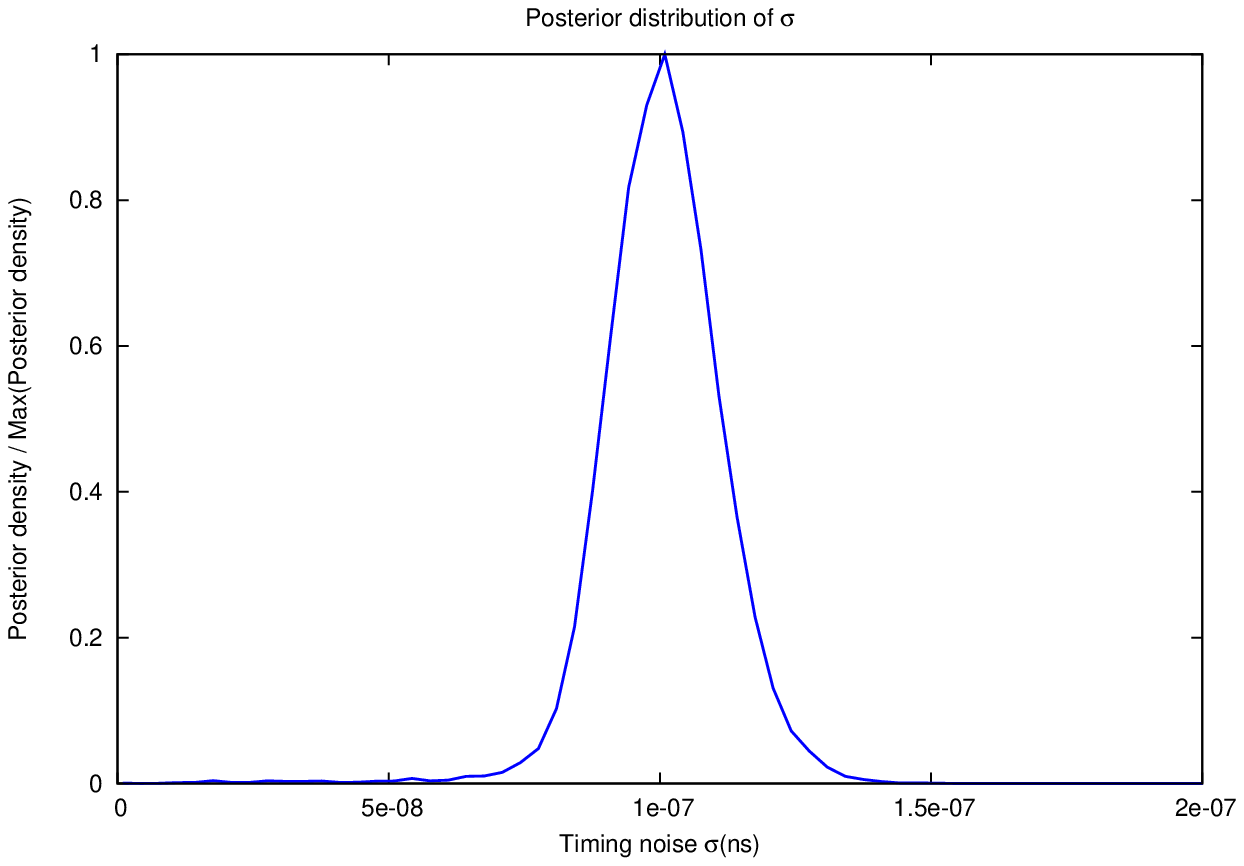}
	\caption{The marginalised posterior of the $a$ parameter, sampled
	  using a Metropolis-Hastings MCMC method.}
	\label{fig:efac}
      \end{figure}

\begin{figure}
	\includegraphics[width=0.5\textwidth]{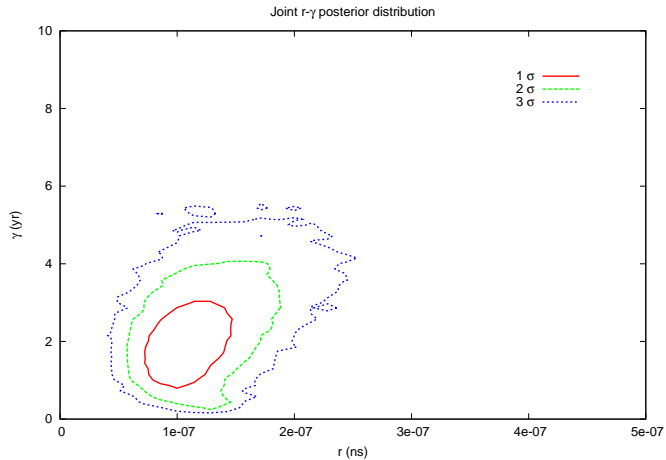}
	\caption{The marginalised posterior of the $r$ and $\gamma$ parameters,
	  sampled using a Metropolis-Hastings MCMC method.}
	\label{fig:rednoise}
      \end{figure}

      Simulating a mock dataset from such a Gaussian distribution is quite
      straightforward; for details see vHLML. We now analyse a mock dataset,
      shown in Figure \ref{fig:residuals}, generated with parameters: $\sigma=100$ns,
      $r=100$ns, and $\gamma=2$yr. We assume uniform prior distributions for all
      parameters: $\sigma \in \left[ 0, 1000 \right]$ns, $r \in \left[ 0, 1000
      \right]$ns, and $\gamma \in \left[ 0, 10 \right]$yr. The posterior is then
      sampled using both MultiNest and a Metropolis-Hastings algorithm, 
      resulting in marginalised posterior distributions as shown in Figure
      \ref{fig:efac} \&\ref{fig:rednoise}. Bayesian evidence values are in good
      agreement between the two methods:
      \begin{eqnarray}
	z_{\text{MCMC}} &=& \exp\left(1523.12 \pm 0.17 \right)  \nonumber \\
	z_{\text{MultiNest}} &=& \exp\left(1522.93 \pm 0.15 \right) .
	\label{eq:residualevidence}
      \end{eqnarray}
      For both methods, the same number of samples has been used: $N=9617$.


    \subsection{Precision tests} \label{sec:errortests}
\begin{figure}
	\includegraphics[width=0.5\textwidth]{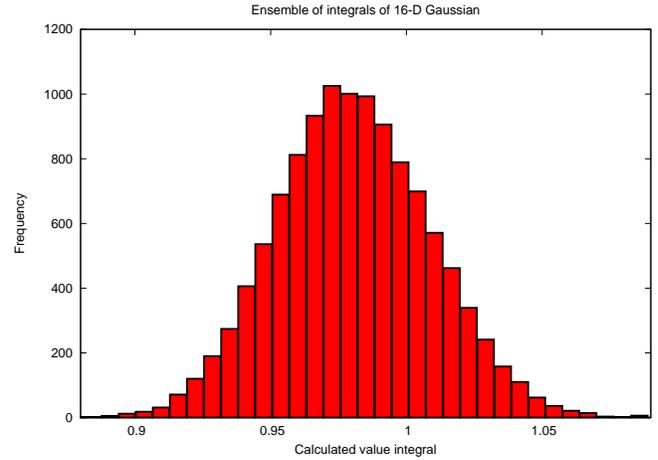}
	\caption{Histogram of the frequency of calculated integral using the
	  method developed in this paper.  We have taken the $16$-dimensional
	  Gaussian of equation (\ref{eq:gaussianmultidimensional}) as integrand.
	  Here we have analysed it with $N=10^5$, and $c=0.3$.  This histogram
	  has mean and standard deviation: $\bar{I} = 0.980$, $\bar{\sigma} =
	  0.028$. The rms of the bootstrap error of the ensemble was
	  $\bar{\sigma}^{\text{BS}}=0.027$}
	\label{fig:hist0}
      \end{figure}

      We now test the accuracy of the algorithm by running it many times on the
      same toy-problem, and then considering the statistics of the ensemble. We
      found the $16$-dimensional Gaussian of section \ref{sec:highdimensional}
      to be an illustrative example. Just as in table \ref{tab:toy1}, we take
      $N=10^5$ and $c=0.3$, and then we run $n=10^4$ Metropolis-Hastings chains on
      this toy-problem. For the $i^{\hbox{th}}$ chain we then calculate the
      integral $I_i$ and bootstrap error estimate  $\sigma^{\text{BS}}$. We have
      presented the results of this analysis as a histogram of $I_i$ values in
      Figure \ref{fig:hist0}. Several useful quantities that characterise the
      ensemble are:
      \begin{eqnarray}
	\bar{I} &=& \frac{1}{n} \sum_{i=1}^{n} I_i = 0.980\nonumber \\
	\bar{\sigma} &=& \sqrt{\frac{1}{n} \sum_{i=1}^{n} \left(I_i -
	\bar{I}\right)^2} = 0.028 \nonumber \\
	\bar{\sigma}^{\text{BS}} &=& \sqrt{\frac{1}{n} \sum_{i=1}^{n}
	\left(\sigma_i^{\text{BS}}\right)^2} = 0.027,
	\label{eq:hist0data}
      \end{eqnarray}
      where $\bar{I}$ is the integral average, $\bar{\sigma}$ is the rms of the
      integral values, and $\bar{\sigma}^{\text{BS}}$ is the rms value of the
      bootstrap errors.

      Figure \ref{fig:hist0} shows that the bootstrap error estimate is quite
      correct, since $\bar{\sigma} \approx \bar{\sigma}^{\text{BS}}$.
      However, though smaller than $\bar{\sigma}$, there is a
      significant deviation in the value of of $\bar{I}$ compared to the true
      value $I=1$.  As we have discussed in section \ref{sec:parametervolume},
      the criterion for the algorithm to converge to the correct value is that
      the small subset $F_{\hbox{t}}$ is sufficiently populated by MCMC samples with
      density proportional to $f(\vec{\Theta})^{-1}$.

      In order to test whether the deviation of $\bar{I}$ from the true value
      $I=1$ is due to
      not fulfilling requirement 2 of section \ref{sec:parametervolume}, we will
      perform 3 additional tests. In all 3 cases, we will construct a new
      ensemble of MCMC chains identical to the ensemble above, except for one
      parameter. The differences with the above mentioned ensemble are:\newline
      1. Instead of a correlated MCMC chain, we use an MCMC chain of
      uncorrelated samples, produced by performing a regular MCMC but only
      storing every $100^{\hbox{th}}$ sample.\newline
      2. $N=10^7$, instead of $N=10^5$. This results in much more samples in the
      subset $F_{\hbox{t}}$.\newline
      3. $c=0.7$, instead of $c=0.3$, which also results in more samples in the
      subset $F_{\hbox{t}}$.

      We present the results of this analysis as the values of equation
      (\ref{eq:hist0data}) in table \ref{tab:histdata}. All adjustments seem to
      improve the precision of the algorithm. Several notes are in
      order:\newline
      1. The only reason we can increase $c$ is because we know exactly what the
      integrand looks like. In practical applications this is probably not an
      option. Also note that bootstrap error increases, indicating that estimate
      of the integral is less stable.\newline
      2. The fact that the uncorrelated chain performs as well as theoretically
      possible according equation (\ref{eq:integralerror}) shows that the algorithm
      suffers significantly under the use of correlated MCMC chains.\newline
      3. Increasing the amount of samples in a chain makes the calculated
      integral more reliable, as the subset $F_{\hbox{t}}$ is more densely populated. Note
      that this large chain is build up of small chains that would yield a
      biased value.
      
      \begin{table}
 	\centering
 	\begin{tabular}{l | r r r r r}
 	  \multicolumn{6}{c}{Ensemble statistics for different parameters} \\
 	  \hline
	  Cor./Unc. & $N$ & $c$ & $\bar{I}$ & $\bar{\sigma}$ & $\bar{\sigma}^{\text{BS}}$ \\
 	  \hline
 	  Cor. & $10^5$ & $0.3$ & $0.980$ & $0.028$ & $0.027$ \\
 	  Unc. & $10^5$ & $0.3$ & $1.000$ & $0.006$ & $0.006$ \\
 	  Cor. & $10^7$ & $0.3$ & $1.000$ & $0.003$ & $0.003$ \\
 	  Cor. & $10^5$ & $0.7$ & $0.994$ & $0.020$ & $0.034$ \\
 	  \hline
 	\end{tabular}
	\caption{The statistics of equation (\ref{eq:hist0data}) for various
	  ensembles of $n=10^4$ MCMC runs on a $16$-dimensional Gaussian. The
	  first chain has the same parameters as used in section
	  \ref{sec:highdimensional}. The other chains differ in either number of
	  samples per chain $N$, the size $c$ of the subset $F_{\hbox{t}}$, or whether or
	  not the samples in a chain are correlated. Note that the error in the
	  uncorrelated chain equals the theoretical Poissonian limit of
	  $\sigma = \sqrt{1 / cN} = 0.006$.}
	\label{tab:histdata}
      \end{table}

  \section{Discussion and conclusions}
    We develop and test a new algorithm that uses traditional MCMC
    methods to accurately evaluate numerical integrals that typically arise when
    evaluating the Bayesian evidence. The new method can be applied to MCMC
    chains that have already been run in the past so that no new samples have to
    be drawn from the integrand. We test the new algorithm on several
    toy-problems, and we compare the results to other algorithms: MultiNest and
    Parallel tempering. We conclude that there is no single algorithm
    that outperforms all others in all cases. High-dimensional, peaked problems
    are better tackled using a traditional MCMC method, whereas very complicated
    multimodal problems are probably best handled with MultiNest. When
    applicable, the new algorithm significantly outperforms other algorithms,
    provided that the MCMC has been properly executed. This new algorithm is
    therefore expected to be useful in astrophysics, cosmology and particle
    physics.

    We have demonstrated that the new algorithm suffers under the use of
    correlated MCMC chains, produced by using the entire chain of a particular
    MCMC method like Metropolis-Hastings. If the new algorithm is used in
    combination with an uncorrelated chain, the accuracy of the numerical
    integral can reach the theoretical limit for stochastic methods: $\sigma =
    I/\sqrt{N}$, with $\sigma$ the uncertainty, $I$ the value of the integral,
    and $N$ the amount of MCMC samples. Using correlated MCMC samples can
    significantly increase the integral uncertainty, and longer MCMC chains are
    needed for the integral to converge. Additional tests to assess convergence
    are required.

    \subsection{Comparison to other work}
      When this paper was already submitted to Monthly Notices, a preprint by
      \citet{Weinberg2009} appeared on the arxiv which also attempts to
      construct an algorithm to calculate the Bayesian evidence using MCMC
      samples of the posterior. Weinberg first discusses the flaws of the
      harmonic mean estimator, and then proposes 2 algorithms to overcome these
      flaws. Weinberg then tests these interesting modifications to the harmonic
      mean estimator on simulated datasets. The algorithm proposed by Weinberg
      bears some resemblance to the algorithm in this paper in that one should
      only use a well-sampled subset of the parameter space when estimating the
      Bayesian evidence.

  \section*{Acknowledgements}
    We would like to thank Yuri Levin, Christian R\"over, Reinhard Prix, Gerard
    Barkema, and Chael Kruip for insightful discussions. This research is
    supported by the Netherlands organisation for Scientific Research (NWO)
    through VIDI grant 639.042.607.


  \bibliographystyle{mn2e.bst}
  \bibliography{vanhaasteren}

\end{document}